\documentclass[fleqn,twoside]{article}
\usepackage{espcrc2}

\usepackage{graphicx}
\usepackage{epsfig}

\newcommand{\be}{\begin{eqnarray}}
\newcommand{\ee}{\end{eqnarray}}
\newcommand{\<}{\langle}
\renewcommand{\>}{\rangle}

\newcommand{\mc}{\mathcal}
\newcommand{\mbf}{\mathbf}
\newcommand{\nn}{\nonumber}

\newcommand{\PR}[4]{{#1}, Phys. Rev. \textbf{#2}, {#3} ({#4})}

\newcommand{\PRL}[4]{{#1}, Phys. Rev. Lett. \textbf{#2}, {#3} ({#4})}

\title{Semileptonic Hyperon Decays on the Lattice: an Exploratory Study}

\author{D. Guadagnoli 
        \address{Dipartimento di Fisica, Universit\`a di Roma ``La Sapienza'', 
                 and INFN, Sezione di Roma, \\
        P.le A.~Moro 2, I-00185 Rome, Italy}, 
        G. Martinelli \addressmark ,
        M. Papinutto \address{NIC/DESY Zeuthen, Platanenallee 6, D-15738
        Zeuthen, Germany} and
        S. Simula \address{INFN, Sezione di Roma Tre, Via della Vasca Navale 84,
                           I-00146, Rome, Italy}
       }
       
\begin{document}

\begin{abstract}
\noindent We present preliminary results of an exploratory lattice study of the vector 
form factor $f_1(q^2 = 0)$ relevant for the semileptonic hyperon decay 
$\Sigma^{-}~\rightarrow~n~l~\nu$. This study is based on the same method used for the 
extraction of $f^+(0)$ for the decay $K^0~\rightarrow~\pi^{-}~l~\nu$. 
The main purpose of this study is to test the method for hyperon form factors in
order to estimate the precision that can be reached and the importance of 
$SU(3)$-breaking effects.
\vspace{1pc}
\end{abstract}

\maketitle

\section{INTRODUCTION}

Semileptonic hyperon decays (SHD) can be considered the ``baryonic way'' to a precise
determination of the CKM matrix entry $V_{us}$ for at least two reasons. First, a recent
phenomenological study \cite{CSW} showed that experimental data on SHDs can be
\emph{combined} to give \emph{direct} access to the quantity 
$~|V_{us}| f_1(q^2 = 0) \equiv |V_{us}| f_1~$ for each decay, since the theoretical input 
of form factors (f.f.) other than  $f_1$ can be neglected, to a very good approximation. 
Inserting the values of $f_1$ predicted by flavor-$SU(3)$ symmetry, the authors of 
Ref.~\cite{CSW} presented an estimate of $|V_{us}|$ for each decay. In this estimate the 
main systematic error comes from assuming $SU(3)$ symmetry for the $f_1$ values.
Although this assumption is reasonable, in view of the Ademollo-Gatto
theorem~\cite{AG}, it should be noted that estimates from various phenomenological 
models~\cite{su3b} predict such corrections at the percent level, thus 
competitive with other corrections as the radiative ones, which are accounted
for in the analysis.

The second reason is that today it is possible to measure f.f.'s with great
accuracy in lattice QCD, thus determining $SU(3)$-breaking corrections in a
model independent way, by use of appropriate double ratios of three-point
functions. This approach was first introduced in Ref.~\cite{fnal} for the study 
of heavy-light f.f.'s and then applied to the $K \rightarrow \pi$ vector f.f. at 
zero momentum transfer (needed in the $K_{l3}$ determination of $V_{us}$) in 
Ref.~\cite{kpi}.

It is then interesting to test the double ratio method on hyperons to see whether 
a comparable precision to that obtained for mesons~\cite{kpi} can be achieved in 
the extraction of $f_1$.

We present the results of a preliminary lattice study of the decay 
$\Sigma^{-}~\rightarrow~n~l~\nu$. We have generated 120 gauge configurations on
a $24^3 \times 56$ lattice at $\beta = 6.20$ ($a^{-1} \simeq 2.6$ GeV), with a
quenched Clover action. We show that the method tested in~\cite{kpi} on mesons 
can be applied to hyperons as well, with results of comparable precision. 

\section{DISCUSSION AND RESULTS}

\begin{figure}[hbt]
\vspace{-0.5cm}
\begin{center}
\includegraphics[scale=0.25]{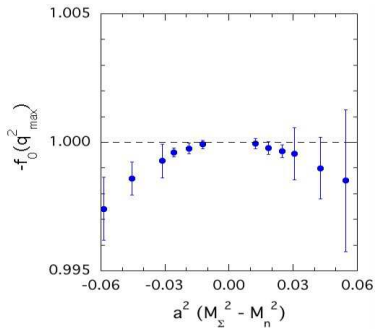}
\end{center}
\vspace{-1.0cm}
\caption{Results for $f_0(q^2_{max})$ vs. the difference of the mass squares of
the external particles.}
\label{fig:f0(qmax)}
\vspace{-0.7cm}
\end{figure}

We are interested in the hadronic matrix element $\mc{M} \equiv \<n| \overline{u} \gamma_\mu
(1 - \gamma_5) s|\Sigma^-\>$ which can be conveniently rewritten in terms of
f.f.'s and external spinors as
\be
\mc{M}& = &\overline{u}_n(p') \left\{ \gamma^\mu f_1(q^2) 
- i \frac{\sigma^{\mu\nu} q_\nu}{M_n + M_\Sigma} f_2(q^2) \right. \nn \\
&&\hspace{-1.5cm}+ \frac{q^\mu}{M_n + M_\Sigma} f_3(q^2) + \left[ \gamma^\mu g_1(q^2) 
- i \frac{\sigma^{\mu\nu} q_\nu}{M_n + M_\Sigma} g_2(q^2) \right. \nn \\
&&\hspace{+0.5cm}+ \left. \left.\frac{q^\mu}{M_n + M_\Sigma} g_3(q^2) \right] 
  \gamma_5 \right\} u_\Sigma(p)
\ee
with $q=p-p'$. One then introduces the ``scalar'' f.f. $f_0(q^2)$ defined as
\be
f_0(q^2) = f_1(q^2) - \frac{q^2}{M_n^2 - M_\Sigma^2} ~ f_3(q^2) ~,
\ee
which at zero $q^2$ coincides with the quantity of interest $f_1(0)$. The main
observation is then that $f_0(q^2)$ can be extracted \emph{well below} $O(1 \%)$ 
accuracy at the kinematical point $q^2_{max}=(M_n - M_\Sigma)^2$ through the 
following double ratio of matrix elements
\be
\displaystyle{\left( \frac{\<n|\overline{u} \gamma_4 s|\Sigma^-\> 
             \<\Sigma^-|\overline{s} \gamma_4 u|n\>}
            {\<n|\overline{u} \gamma_4 u|n\> 
             \<\Sigma^-|\overline{s} \gamma_4 s|\Sigma^-\>}
\right)_{\mbf{p}=\mbf{p'}=0}} = 
[f_0(q^2_{max})]^2~,\nn
\ee
\vspace{-0.35cm}
\be
\label{ratio}
\ee
where both external particles are at rest. The advantages of the ratio (\ref{ratio}) 
are described in Ref.~\cite{kpi} and thus not reported here. The values of $q^2_{max}$ 
can be taken \emph{very close} to zero [$O(10^{-3})$ in units of the lattice spacing] 
by appropriately choosing the values of the hopping parameter of the simulation. 
We chose $k \in \{0.1336, 0.1340, 0.1343, 0.1345\}$ corresponding to pseudoscalar (PS) 
masses in the range $1.1$ GeV $\geq M_{PS} \geq$ $0.75$ GeV. The results for $f_0(q^2_{max})$
are reported in Fig.~\ref{fig:f0(qmax)}. From the y-axis scale one can
appreciate the high precision obtained in the values of $f_0$.

To extrapolate $f_0(q^2)$ to $q^2 = 0$ one has to use values of $f_0(q^2)$ and $f_1(q^2)$ 
obtained through usual f.f. analysis. Given the high accuracy
by which the point at $q^2_{max}$ is determined, and its closeness to $q^2=0$, it
is enough to reach for $f_{0,1}(q^2)$ an accuracy of $O(10-20 \%)$. Similarly to the case of
mesons~\cite{kpi}, $f_1$ turns out to be quite well determined, while for $f_0$ one has to 
resort to other appropriate double ratios, which give access to the quantities $f_0(q^2)/f_1(q^2)$. 

\begin{figure}[h!]
\vspace{-0.5cm}
\begin{center}
\includegraphics[scale=0.25]{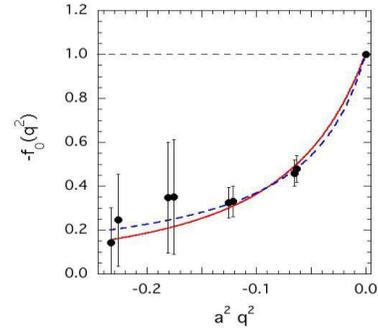}
\end{center}
\vspace{-1cm}
\caption{Results for $f_0(q^2)$ vs. $a^2 q^2$. The dashed and solid lines represent a monopole and a
dipole fit to the data, respectively [see Eq.~(\ref{curves})].}
\label{fig:f0(q2)}
\vspace{-0.5cm}
\end{figure}

\begin{figure}[h!]
\vspace{-0.5cm}
\begin{center}
\includegraphics[scale=0.25]{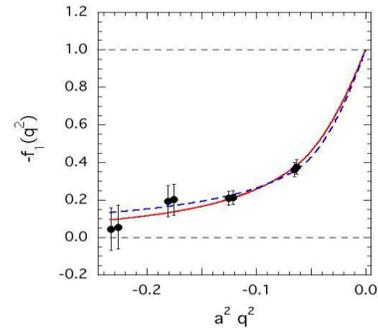}
\end{center}
\vspace{-1cm}
\caption{Same as Fig.~\ref{fig:f0(q2)}, but for $f_1(q^2)$.}
\label{fig:f1(q2)}
\vspace{-0.7cm}
\end{figure}

In Figs.~\ref{fig:f0(q2)} and \ref{fig:f1(q2)} we display the results for $f_0(q^2)$ and $f_1(q^2)$
respectively, for a specific combination of quark masses, i.e.
$M_\Sigma(0.1345,0.1343)$ and $M_n(0.1345)$. 
The points are paired due to the fact that both $\Sigma^-
\rightarrow n$ and $n \rightarrow \Sigma^-$ were considered in the analysis. From
Fig.~\ref{fig:f0(q2)} one can also contrast the remarkable accuracy of the point
at $q^2_{max}$ (the rightmost one) versus the other $q^2$ 
values. 

We then extrapolated to the value $f_0(0) = f_1(0)$ by fitting the data in
Figs.~\ref{fig:f0(q2)} and~\ref{fig:f1(q2)} with appropriate model functions,
i.e. 
\be
F_m(x)&=& \frac{A}{1-\frac{x}{B}} \qquad \mbox{monopole fit},\nn \\
F_d(x)&=& \frac{C}{\left(1-\frac{x}{D}\right)^2} \qquad \mbox{dipole fit},
\label{curves}
\ee
also shown in Figs.~\ref{fig:f0(q2)} and~\ref{fig:f1(q2)} as dashed and solid
lines, respectively. Both model functions in~(\ref{curves}) turn out to
describe well the data, and the dipole fit parameter $\sqrt{D}$ agrees with the
value predicted by pole dominance (that is, the $K^*$ meson mass) within 15 \%
accuracy.

\begin{figure}[hbt]
\vspace{-0.5cm}
\begin{center}
\includegraphics[scale=0.25]{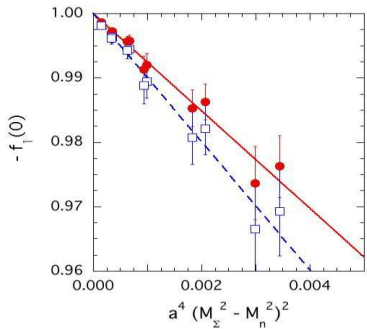}
\end{center}
\vspace{-1cm}
\caption{Results for $f_1(0)$ vs. the difference $a^4
(M_\Sigma^2 - M_n^2)^2$, for data extrapolated through a monopole (open squares)
and a dipole fit (full circles). The dashed and solid lines represent linear
fits of the data themselves, according to the Ademollo-Gatto theorem.}
\label{fig:f1(0)_m}
\vspace{-0.5cm}
\end{figure}

\begin{figure}[hbt]
\vspace{-0.8cm}
\begin{center}
\includegraphics[scale=0.25]{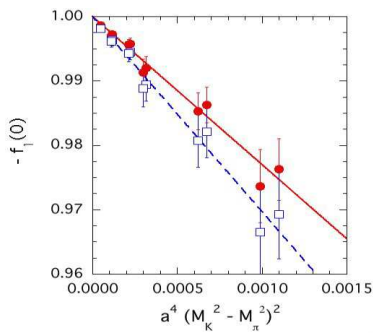}
\end{center}
\vspace{-1cm}
\caption{Same as Fig.~\ref{fig:f1(0)_m}, but for the x-axis quantity $a^4
(M_K^2 - M_\pi^2)^2$.}
\label{fig:f1(0)_b}
\vspace{-0.5cm}
\end{figure}

We finally show the results for the extrapolated values of $f_1(q^2 = 0)$ at
the different masses used in the simulation. They are plotted in
Figs.~\ref{fig:f1(0)_m} and~\ref{fig:f1(0)_b} versus the mass differences $a^4
(M_\Sigma^2 - M_n^2)^2$ and $a^4 (M_K^2 - M_\pi^2)^2$, respectively, both of which behave as
the square of the $SU(3)$-breaking parameter $m_s - \overline{m}$, $m_s$ being the
mass of the strange quark, and $\overline{m}$ the common mass of the
$u,d$ quarks. Open squares and full circles refer to the data points extrapolated 
through a monopole and a dipole fit [see Eq.~(\ref{curves})], respectively.

Both figures display a nice linear dependence, as expected from the Ademollo-Gatto
theorem. This behavior was checked through a linear fit, also shown in 
Figs.~\ref{fig:f1(0)_m} and~\ref{fig:f1(0)_b} as dashed lines for monopole data
and solid lines for dipole data. We notice that the $SU(3)$-breaking effect 
is resolved as a function of the hadron masses with good precision even with our 
limited statistics.

In conclusion, this study shows that a remarkable precision (comparable to that of
$K-\pi$~\cite{kpi}) can be achieved for the $SU(3)$-breaking quantity $f_1(0) -
f_1(0)_{SU(3)}$ for hyperons as well. The crucial issue is then the accuracy in
the extrapolation to the physical point, for which it is essential to understand 
the dependence on the quark masses and the role played by ChPT. An extensive lattice 
study of all the (non-isospin equivalent) SHDs will then follow.

\end{document}